# Configuration and Collection Factors for Side-Channel Disassembly

Random Gwinn
*Johns Hopkins University*
*Applied Physics Laboratory*
Laurel, MD, USA
random.gwinn@jhuapl.edu

Mark Matties
*Johns Hopkins University*
*Applied Physics Laboratory*
Laurel, MD, USA
mark.matties@jhuapl.edu

Aviel D. Rubin
*Johns Hopkins University*
*Whiting School of Engineering*
Baltimore, MD, USA
rubin@jhu.edu

*Abstract*— **Myriad uses, methodologies, and channels have been explored for side-channel analysis. However, specific implementation considerations are often unpublished. This paper explores select test configuration and collection parameters, such as input voltage, shunt resistance, sample rate, and microcontroller clock frequency, along with their impact on side-channel analysis performance. The analysis use case considered is instruction disassembly and classification using the microcontroller power side-channel. An ATmega328P microcontroller and a subset of the AVR instruction set are used in the experiments as the Device Under Test (DUT). A time-series convolutional neural network (CNN) is used to evaluate classification performance at clock-cycle fidelity.**

**We conclude that configuration and collection parameters have a meaningful impact on performance, especially where the instruction-trace's signal to noise ratio (SNR) is impacted. Additionally, data collection and analysis well above the Nyquist rate is required for side-channel disassembly. We also found that 7V input voltage with 1 k$\Omega$ shunt and a sample rate of 250 – 500 MSa/s provided optimal performance in our application, with diminishing returns or in some cases degradation at higher levels.**

## I. Introduction

The foundation of contemporary side-channel analysis is often credited to Kocher [1], [2]. Since then, the research area has grown significantly beyond its cryptanalysis roots. Side-channel disassembly, whereby the specific instructions executed on a device, sequence order, and other details can be deduced through monitoring and analysis of side-channel emissions, is one such example. This can be a powerful tool for non-invasive reverse engineering of embedded code at runtime, evaluating information leakage of devices, or even using the analog channel as a means of device monitoring for cross-domain security. However, current literature lacks implementation details necessary for reproducibility and a methodical exploration of parameters that could inform technique optimizations and transition of research to practice.

Table I shows several power side-channel works and the parameters employed. Sample rate and shunt resistance were either constants or not listed at all. Goldack [3] was the only work that explicitly varied DUT frequencies, but only did so as an initial exploration without quantified performance assessment. Besides [3] and [4], input voltage was not listed, and no works assessed its impact on performance.

Contributions of this work include controlled experimentation and analysis of: input voltage; shunt resistance; trace amplitude and standard deviation (STD); sample rate, microcontroller clock frequency, and microcontroller clock frequency to sample rate ratio; as they relate to power side-channel disassembly classification accuracy. These factors would also impact

TABLE I. Literature Parameters

| Work | Sample Rate (MSa/s) | DUT Freq (MHz) | Shunt ($\Omega$) |
|---|---|---|---|
| VWG.2007 [5] | ? (a) | 4 | Not listed |
| G.2008 [3] | Not listed | 1 (b) | Not listed |
| EPW.2010 [6] | 1,000 | 1 | Not listed |
| MMM.2014 [4] | 500 | 4 | ? (c) |
| LWZ.2016 [7] | 1,250 | 11.0592 | 46.7 (d) |
| PT.2017 [8] | 1,250 | 16 | Not listed |
| PXJ.2018 [9] | 2,500 | 16 | 330 |
| KLR.2020 [10] | 500 | 10 | 1 (e) |

a. Unclear from description; author seems to conflate bandwidth with sample rate and describes resampling to DUT frequency.
b. Partial exploration of 32 kHz and 4 MHz also indicated.
c. Unclear from description.
d. Shunt on supply side.
e. Measurement was across 4-separate shunts on each supply pin with amplification.

other side-channel analysis areas, but may do so in different ways; therefore side-channel disassembly (with defined DUT, environment, and analysis approach) was selected to serve as the common reference point for measuring the impact of the considered configuration and collection parameters. Further work from the community would be necessary to generalize beyond these conditions. Additionally, several areas that would benefit from additional research have been identified.

This paper is organized with an overview of the evaluation approach and experimental setup in section II; a description of conducted experiments and results in section III; and conclusions in section IV.

## II. Evaluation Approach

This paper leverages the setup, analysis approach, and concluded employment used in [11], while instead exploring input and collection variables as the research focus. Given that, specific details of side-channel analysis, disassembly, machine learning, or other topics are beyond the primary scope of this paper.

The fixed setup, as described in sub-section II.A, remained constant across experiments within this paper. The factors that were varied in support of experimentation are described in sub-section II.B.

### A. Fixed Experimental Setup and Baseline Analysis

The DUT in this paper was an Elegoo Uno R3 [12] with an ATmega328P [13] microcontroller. Power side-channel traces were collected from the ground-side of the microcontroller using a digital sampling oscilloscope with 200 MHz bandwidth at a memory depth of 14 Mpts. A custom program loop comprised of a balanced-pairing of instructions from the AVR [14] subset listed in Table II was executed on the DUT.

Program loop, instruction, and clock-cycle instances were extracted from aligned oscilloscope traces files. Extracted data was labeled and transformed into scalograms, then time-sequenced for input to the time-series CNN. No other filtering or signal amplification was employed. Analysis was performed offline in Python 3 notebooks with the classifier built with TensorFlow [15] and Keras [16]. For each experiment, 70% of provided data was used for training while 30% was held-out as the test set.

### B. Varied Experimental Setup

In addition to the fixed setup described above, Table III lists the factors that were deliberately varied in this research. Additional details along with experimental results are described in section III.

TABLE II. AVR INSTRUCTION SUBSET

| Instruction | Clock-Length | Type |
|---|---|---|
| sbi | 2 | bit / IO |
| nop | 1 | control |
| add | 1 | arithmetic |
| sub | 1 | arithmetic |
| cbi | 2 | bit / IO |
| push | 2 | transfer |
| pop | 2 | transfer |
| mul | 2 | arithmetic |
| eor | 1 | logic |
| movw | 1 | transfer |
| rjmp | 2 | program flow |

TABLE III. SUMMARY OF VARIABLES EXPLORED

| Factor varied | Values considered |
|---|---|
| DUT Input Voltage (V) | 5, **7***, 9 |
| Shunt resistance ($\Omega$) | 0, 1, 2.7, 6.8, 10, 100, 330, **1k***, 4.7k, 8.2k, 10k, 12k, 15k, 33k, 100k, 1M |
| Oscilloscope Sample Rate (MSa/s) | 5, 10, 20, 50, 100, 250, **500***, 1000 |
| Subsample Rate (MSa/s) | 5, 10, 20, 50, 100, 250, 500 |
| Microcontroller Clock-Frequency (Hz) | **1M***, 2M, 4M, 8M, 16M |
| Training/Test Data quantity | **Fixed loops***, fixed data |

*\* denotes default values that were used within this work unless varied for experimentation purposes as indicated*

## III. Experiments and Results

### A. Input Voltage

The DUT has a manufacturer stated acceptable voltage input range from 7V to 12V [12] in its standard configuration with the ATmega328P operating at 16 MHz. The DUT can, however, operate with lower input voltage, especially with reduced clock-speed and settings. In this experiment, the DUT is operated at 1 MHz with low-power settings. At each input voltage, 225 program loops were assessed. The experiment was conducted 10 times, with 125 epochs each.

As Fig. 1 shows, 7V provided the best classification performance and a basis to compare the 5V and 9V cases. Although the 5V case exists outside of the specified (default) voltage input range, the DUT executes the program loop without issue, and a clear side-channel trace is obtained. However, a significant drop in classification performance is observed, which is attributed to the reduced amplitude of the side-channel trace. Additionally, in the 9V case increased side-channel trace amplitude is not realized due to power regulators within the DUT. A question remains why the 9V case produced lower classification performance, and we hypothesize that the increased activity from the DUT power regulators may have introduced additional noise to the obtained side-channel traces. Future research is required to explore this and other input voltage considerations.

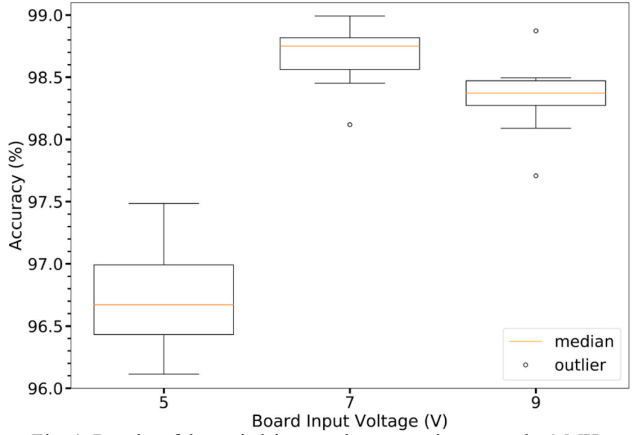

Fig. 1: Results of the varied input voltage experiment on the 1 MHz DUT are shown with boxplots indicating the distribution of results over the 10 conducted trials. The 5V case provided the lowest accuracy with highest variance of test cases. 7V performed best, while 9V had a slight reduction in accuracy that we suspect is caused by increased noise from the DUT's voltage regulator.

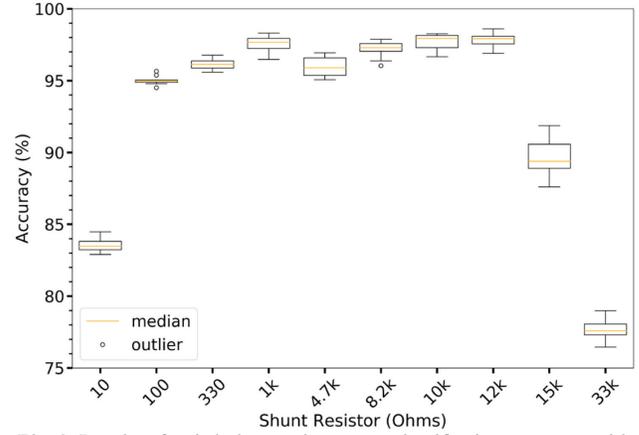

Fig. 2: Results of varied shunt resistance on classification accuracy with DUT operating at 1 MHz. Distribution of results over 10 trials depicted as boxplots shows 1kΩ being optimal in our application. Observed performance increases until 1kΩ, with plateau and then degradation as resistance increases further.

*B. Shunt Resistance*

Power-side channel traces are obtained by measuring voltage drop across a shunt resistor with known resistance. Given this, the impact of the precision and ohmic value of employed shunts are important factors to consider. In this research, all shunts used were rated with 1% precision, and verified to be within specified tolerances using a digital multimeter. Rated resistance values from 0 Ω to 1 MΩ were explored.

In this experiment, 80 program loops at each shunt value was used for classification assessment. The experiment was conducted 10 times, with 125 epochs each, and results are shown in Fig. 2. Although side-channel trace data was collected for shunt values less than 10 Ω, these trace-sets produced low SNR and inconsistently produced clear signatures. Therefore the <10 Ω data was excluded in this analysis, while refinements to setup or analysis may exist to make low ohm shunts usable (e.g., amplifiers, filtering, boosting, etc.).

Similar to the low ohm case, high ohm datasets did not provide clear results. The 1 MΩ data was entirely excluded as in the <10 Ω case. The 33 kΩ and 100 kΩ datasets were able to be partially extracted and assessed, but suffered from significant variation. Given the intent to explore classification performance across as wide a resistance range as possible, the amount of data used within this experiment (80 program loops per shunt value) was driven by the maximum number of usable loops from the 33 kΩ and 100 kΩ datasets. With less training data, the absolute classification accuracy is reduced slightly from other experiments, but performance relative to resistance is obtained.

As shown in Fig. 2, accuracy increases with resistance before plateauing from 1 kΩ to 12 kΩ, then ultimately decreasing. The 1 kΩ shunt, therefore, provides the highest direct classification performance with least amount of circuit-loading. We also examined 100 kΩ, but the accuracy was quite low (approximately 25%) and had high variation. Thus, we omitted it from Fig. 2.

*C. Trace Amplitude Analysis*

Leveraging the data from the shunt ohms experiment in section III.B., the amplitude range (peak to trough) of individual clock-cycle instances throughout the executed program loop were explored. The clock-cycle instance with maximum, minimum, and average amplitude ranges across all instructions within the program loop were calculated for each shunt value as depicted in Fig. 3. The results show a consistent profile, with amplitude range relative to shunt resistance in the min, max, and average cases. Additionally, the normalized STD of the average amplitude range for each shunt value at each program loop instruction instance was calculated and depicted as the secondary y-axis in Fig. 3. The normalized STD has a nearly-inverse relationship to the average amplitude range. The 1 kΩ dataset produces the highest amplitude with the least variance.

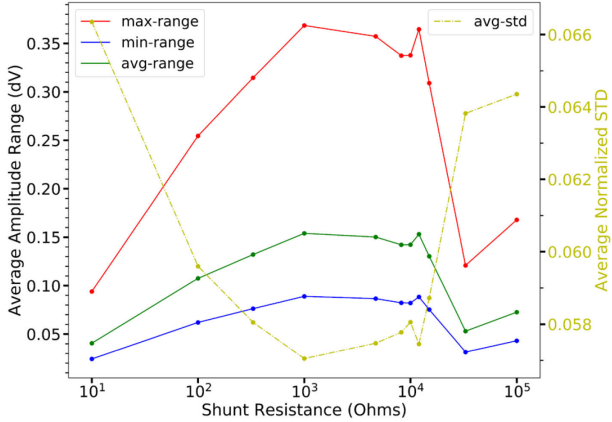

Fig. 3: Averaged amplitude range of program loop clock-cycle indices, depicting the minimum, maximum, and average case across different shunt resistances with the DUT operating at 1 MHz. Normalized STD also shown, with a nearly inverse relationship to amplitude.

Exploring the relationship between amplitude range and classification performance further, averages from the input voltage experiment (section III.A.), and a subset of averages from the shunt resistance experiment (section III.B.) were analyzed as depicted in Fig. 4. From this, a positive correlation is found between amplitude and classification accuracy. After calculation, the Pearson correlation coefficient was identified as r = 0.91. Note: shunt values at and above 12 kΩ were excluded as they exist as outliers to the data beyond what could be readily explained by their increased STD as previously noted. Future analysis is necessary to explore root cause.

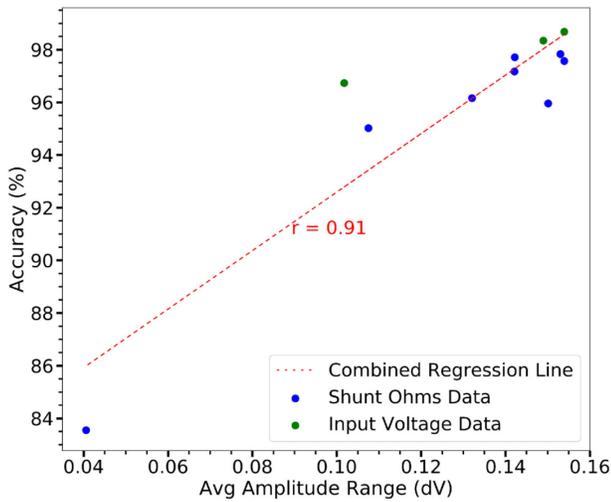

Fig. 4: Analysis of the shunt ohms and input voltage datasets shows a strong positive correlation between trace amplitude and instruction classification accuracy.

### D. Sample Rate

The oscilloscope sample rate used to capture side-channel traces is thought to be a critical factor in analysis outcomes, as it provides the temporal resolution. At the most basic level, one might assume sampling at the Nyquist rate would be sufficient. Conversely, collecting at arbitrarily high rates may present an arduous amount of data, including unnecessary higher frequency components, which could degrade performance. This experiment seeks to quantify the impact of sample rate on achieved performance to inform proper selection.

In the evaluation of sample rates, analysis was performed using either a fixed number of program loops (Fixed Loops), or using a fixed amount of raw trace data (Fixed Data). The oscilloscope was configured with a fixed memory depth, where all collections occur at 14 Mpts per collected trace file. At lower sample rates, individual program loop executions are captured and represented within the trace data by a lower number of discrete time samples, whereas each trace file consists of a greater number of program loops. As sample rate increases, so do the number of time samples per program loop, thus reducing the number of captured program loops per trace file.

In the fixed loop case, 45 program loops from each of 5 captured trace files are extracted. The number of program loops was selected based on the 1 GSa/s dataset as the limiting case, where 48 whole program loops are the maximum possible at current conditions. Although 45 loops is somewhat arbitrary, it does make use of more than 92% of captured data, along with the necessary margin for alignment and extraction of program loops from raw trace files. The executed program loop is 287 clock-cycles in length, with each clock-cycle instance extracted as training and test data. This results in 64,575 clock-cycle samples used in all fixed loop assessments. Table IV lists the program loop and clock-cycle quantities for fixed data assessments.

An experiment was conducted on the DUT with a 1 MHz microcontroller clock frequency to assess the impact of sample rate on classification performance, with results shown in Fig. 5. The experiment was repeated 10 times, with 125 epochs each. Although 1 GSa/s achieves the highest accuracy, diminishing returns are observed vs. the 500 MSa/s dataset. Also, as expected, providing more training data in the fixed data case increases performance over that of the fixed loop case. However, even a significant increase in training data does not overcome the decrease in performance at lower sample rates. Not shown in Fig. 5 is 50 MSa/s, which produced an average fixed loop classification accuracy of 81%. Fixed data assessment of 50 MSa/s was not conducted due to the considerable training time involved given the number of clock-cycle samples. We believe that the 50 MSa/s fixed data results would follow a similar trend to the results from other test cases in this experiment.

TABLE IV. FIXED DATA DESCRIPTIONS

| Rate (MSa/s) | # Program Loops | # Clock-cycle Samples |
|---|---|---|
| 50 | 4,500 | 1,291,500 |
| 100 | 2,250 | 645,750 |
| 250 | 900 | 258,300 |
| 500 | 450 | 129,150 |
| 1000 | 225 | 64,575 |

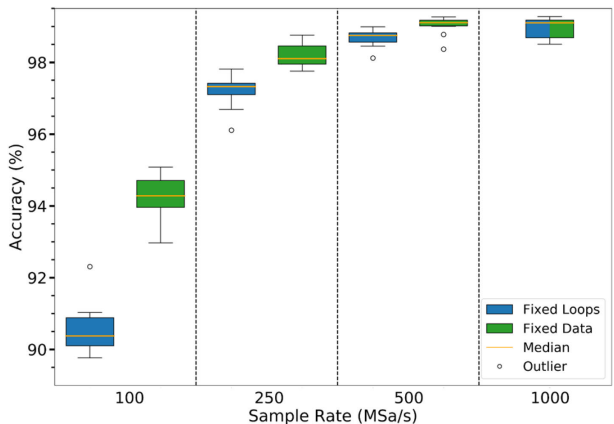

Fig. 5: Results of the varied sample rate experiment on the 1 MHz DUT are shown with boxplots indicating the distribution of results across 10 trials. Separate boxplots are provided for the fixed loop and fixed data cases described in section III.D. The fixed loop case is driven by the 1000 MSa/s quantity, therefore at this sample rate the fixed loop and fixed data cases are the same. As expected, additional data in the fixed data cases produce improved results.

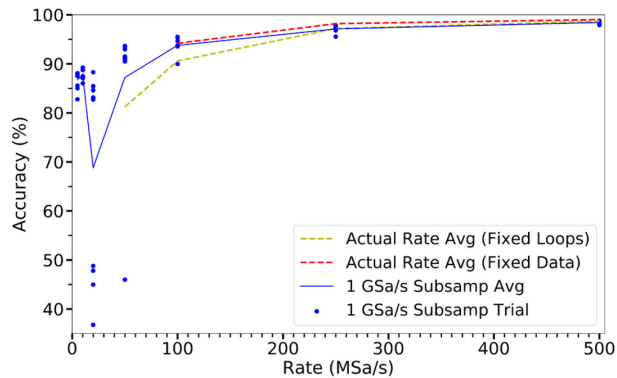

Fig. 6: Classification performance of subsampled data across 10 trials shown as the scatterplot. Trace data collected at 1 GSa/s with the DUT operating at 1 MHz. Classification performance of data collected at actual rates shown as dashed lines for comparison. Results generally match with decreased accuracy and higher variance at lower rates.

### E. Subsample Rate

To supplement the actual oscilloscope sample rate captures described in section III.D., data from the 1 GSa/s collection was used to produce subsampled datasets. These datasets contain time samples per program loop and clock-cycle that correspond to 500, 250, 100, 50, 20, 10, and 5 MSa/s. The subsampled data was then assessed and compared to actual oscilloscope sample rate results where available.

In this experiment, 225 program loops at each sample and subsample rate was assessed. The experiment was conducted 10 times, with 125 epochs each. For the subsampled data, each of the 10 trials are depicted as a scatter plot and average line in Fig. 6. Additionally, the average of 10 trials at each actual sample rate for both the fixed loops and fixed data cases, per section III.D., are overlaid to provide a basis of comparison.

As shown in Fig. 6, the classification performance results of actual sample rate and subsampled data are largely consistent. This is true of the average trend across sample rates as well as the increased variation at specific rates, especially at 20 and 50 MSa/s. It is interesting that variation does not increase consistently as sample rate decreases, but instead seems to impact certain rates more than others, while very low rates such as 5 and 10 MSa/s have variation similar to that of 100 MSa/s. We believe that this may be due to periodic mis-alignment of key trace features at certain rates (similar to a harmonic). Below these rates key features are maintained, but with less fidelity, and above these rates the capture ratio to individual clock-cycles is large enough to avoid impact from the mis-alignment.

Additionally, subsampled sets were also created and assessed from 500 MSa/s capture source data. Subsample assessment of the 500 MSa/s source was largely consistent to the 1 GSa/s source, with a slight reduction in classification performance on average. Although subsampled data cannot completely replace actual sample rate data, consistency between subsample assessments and actual collection rate data allows additional and more expeditious experimentation given the time, storage, and processing requirements involved with collecting and assessing multiple distinct sets. Subsample results of interest can then inform targeted data collection and assessment activities to confirm the findings with actual data. Future work may explore interpolation as a potential means of increasing classification performance of under-sampled data.

### F. Microcontroller Clock Frequency

Another factor we believed to be key in side-channel analysis outcomes is the operating frequency of the DUT's microcontroller. Further, based upon observations from the 1 MHz analyses, we hypothesize that the ratio between microcontroller frequency and sample rate is actually the important consideration rather than microcontroller frequency or sample rate alone. An experiment was conducted to explore this hypothesis by assessing the classification performance across varied microcontroller clock frequencies and sample rates.

Operating at 1 MHz (using the DUT's internal 8 MHz clock and divider), trace data was captured at 50, 100, 250, 500, and 1000 MSa/s. Operating at 4 MHz (using an external crystal oscillator), trace data was captured at 100, 250, 500, and 1000 MSa/s. Operating at 8 MHz (using the DUT's internal 8 MHz clock without divider), trace data was captured at 250, 500, and 1000 MSa/s. Program loops were not able to be readily extracted at lower sample rates and higher clock frequencies, so common results across operating frequencies are limited to 250, 500, and 1000 MSa/s. In all cases, 5 trace files were captured at 14 Mpts for each operating frequency and sample rate combination.

The experiment then assessed 225 program loops from each test case across 10 trials of 125 epochs each. The results are shown in Fig. 7, with (a) showing the results by sample rate while (b) shows the results by the ratio of sample rate to the microcontroller clock frequency of the DUT. Surprisingly, the 4 MHz and 8 MHz cases outperform the hypothesized relationship and in some cases achieve greater overall results than the 1 MHz base case. There does, however, appear to be a case-specific performance curve rather than following the expected results from the 1 MHz case. It is also interesting that in the 4 MHz case, performance actually decreases slightly at higher sample rates and ratios. Initial speculation was that abnormally low performance in one or more trials may have brought-down the average. However, examination of individual trials data shows this not to be the case. Within each of the microcontroller frequency datasets, the highest observed STD occurred at the lowest sample rate and ratio. In the 4 MHz case, the minimum STD occurred at 250 MSa/s (also its peak accuracy), but only a slight increase in STD was observed at higher sample rates (remaining amongst the lowest STD values globally), which would not explain the reduced accuracy. Another question, is whether this is similar to the 1 MHz 20 and 50 MSa/s cases, with localized performance degradation as compared to the trend. Additional research is required to explore these results further.

Questions arising from higher than expected performance, and significant visual differences in clock-cycle traces at different microcontroller clock frequencies, led to exploratory data collections with the DUT operating at 2 MHz and 16 MHz. Fig. 8 shows a clock-cycle trace of the *eor* instruction at index 283 of the program loop. The trace is averaged across 10 program loop instances at 1 GSa/s for each of the collected DUT operating frequencies. This data is then subsampled to represent visual trace changes at reduced rates.

As previously noted by Goldack [3], the distance between side-channel trace peaks decreases at increased clock frequency. This effect is most clear with the 1 MHz and 2 MHz cases shown in Fig. 8: the trace signature is quite similar but with reduced flat-space in the center between peaks. However, at 4 MHz and higher, no flat-space exists and the overall trace signature changes significantly. In part, as Goldack stated [3], residual capacitance is likely contributing; however, we also believe that at specified rates these traces are under-resolved and unable to show subtleties in the trace signature.

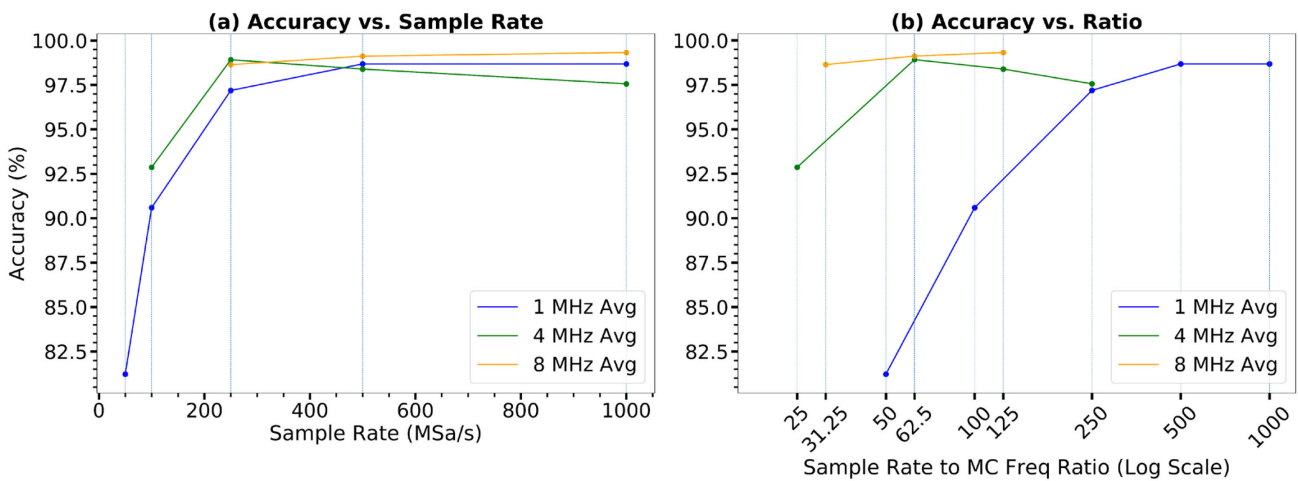

Fig. 7: Results of the varied DUT operating frequency experiment: (a) shows the classification accuracy on the test set for the DUT operating at 1, 4, and 8 MHz and collected by the oscilloscope between 50 – 1000 MSa/s; (b) shows the same results organized by sample rate to operating frequency ratio rather than sample rate. Contrary to our hypothesis, ratio does not generally inform accuracy as the 4 and 8 MHz cases achieve similar performance to 1 MHz at the same sample rates and lower ratios. Instead, the results indicate each operating frequency may have its own rate/ratio to accuracy curve or additional unknown factors are influencing obtained results.

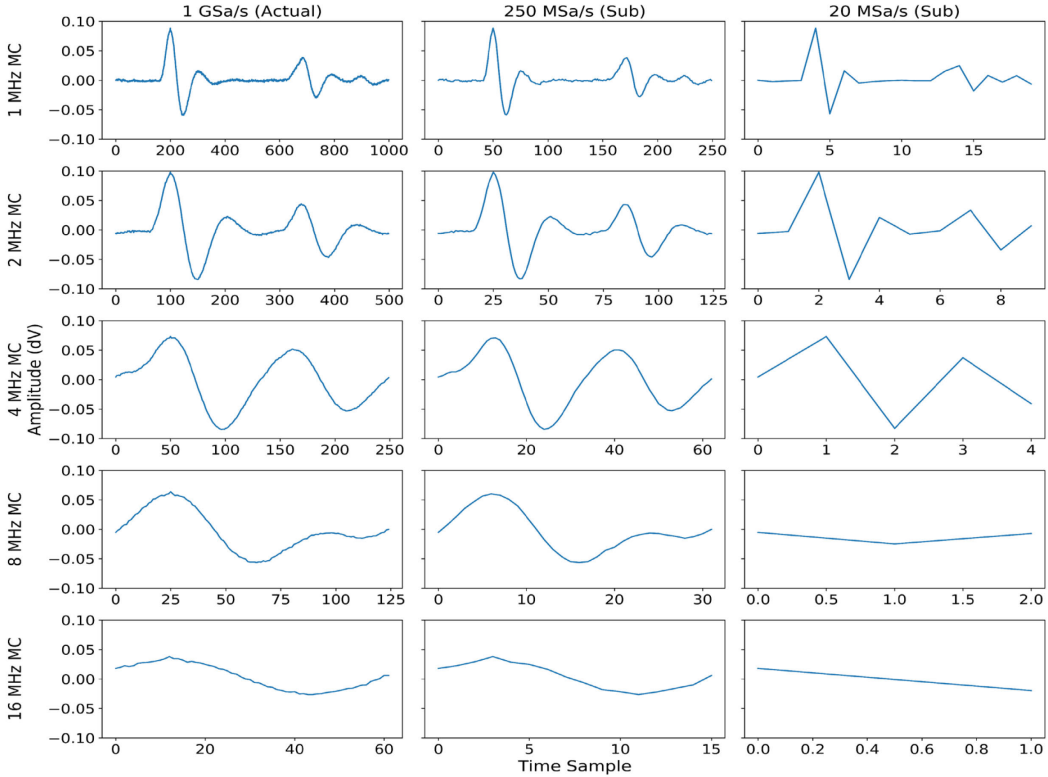

Fig. 8: Example averaged trace of the *eor* instruction at program loop clock-cycle index 283 executed with the DUT operating at 1, 2, 4, 8, and 16 MHz; as captured by the oscilloscope at 1 GSa/s and subsampled to 250 and 20 MSa/s. Each instruction instance exhibits a unique trace signature due in part to the pipelined execution. Visual details of this signature are lost at lower sample rate to operating frequency ratios. Each plot's sample rate to microcontroller frequency ratio is the same as its number of time samples per clock-cycle on its x-axis. Relative distance between amplitude peaks decreases at higher operating frequencies, regardless of sample rate or ratio.

The higher than expected performance of the 4 MHz and 8 MHz datasets, despite their traces represented nearly as varied-amplitude sinusoids, is particularly interesting. During analysis of the 1 MHz dataset, it was believed that the characterization of side-channel traces with multiple localized major/minor ripples and some distinct shape differences was critical to instruction and clock-cycle discrimination. However, given these new results it may be possible to reduce complexity of the classification approach. Additional research is required to explore these results and alternate techniques further.

## IV. CONCLUSIONS

In this work we have assessed select side-channel analysis test configuration and collection parameters which are poorly-defined and under-explored within current literature. Factors such as input voltage, shunt resistance, sample rate, and microcontroller clock frequency, were examined for their impact on side-channel analysis performance.

We have shown that test configuration and collection parameters can contribute significantly to analysis outcomes, especially where SNR is impacted. Collection at twice the DUT operating frequency, as informed by the Nyquist rate, would prove insufficient for side-channel disassembly based upon analysis thus far; however, collection at arbitrarily high rates is unnecessary. We found that 7V input voltage with 1 k$\Omega$ shunt and a sample rate of 250 – 500 MSa/s provided optimal performance in our application (when balanced against other trades), with diminishing returns or in some cases degradation at higher levels.

Configuration and collection parameters in this work were assessed for the specific use case of power side-channel disassembly with a single DUT and classification approach. Therefore, specific findings should not be assumed to generalize without further work. However, we believe the broader factors will extend to other applications and environments. Researchers and developers should therefore consider these factors in their environment, and in some cases tuning them just as they would the architecture, parameters, and hyper-parameters of their machine learning models to achieve optimal performance given their resources and objectives.

## REFERENCES

[1] P. Kocher. (1996). "Timing Attacks on Implementations of Diffie-Hellman, RSA, DSS, and Other Systems." Presented at the 16th Annual International Cryptology Conference on Advances in Cryptology. [Online]. Available: https://paulkocher.com/doc/TimingAttacks.pdf

[2] P. Kocher, J. Jaffee, and B. Jun, "Differential power analysis." Proceedings of the 19th Annual International Cryptology Conference on Advances in Cryptology, Springer-Verlag, 1999, pp. 388-397.

[3] M. Goldack, "Side-Channel based Reverse Engineering for Microcontrollers," Thesis, Ruhr-University Bochum, 2008.

[4] M. Msgna, K. Markantonakis, and K. Mayes. (2014). "Precise instruction-level side channel profiling of embedded processors." Proceedings of the 10th International Conference on Information Security Practice and Experience - Volume 8434. Fuzhou, China, Springer-Verlag: 129–143.

[5] D. Vermoen, M. Witteman, and G. Gaydadjiev. (2007). "Reverse Engineering Java Card Applets Using Power Analysis." Information Security Theory and Practices. Smart Cards, Mobile and Ubiquitous Computing Systems, Berlin, Heidelberg, Springer Berlin Heidelberg.

[6] T. Eisenbarth, C. Paar, and B. Weghenkel. "Building a side channel based disassembler." In: Gavrilova M.L., Tan C.J.K., Moreno E.D. (eds) Transactions on Computational Science X. Lecture Notes in Computer Science, vol 6340. Springer, Berlin, Heidelberg. https://doi.org/10.1007/978-3-642-17499-5_4

[7] Y. Liu, L. Wei, Z. Zhou, K. Zhang, W. Xu, and Q. Xu. (2016). "On Code Execution Tracking via Power Side-Channel." Proceedings of the 2016 ACM SIGSAC Conference on Computer and Communications Security. Vienna, Austria, Association for Computing Machinery: 1019–1031.

[8] J. Park and A. Tyagi. (2017). "Using Power Clues to Hack IoT Devices: The power side channel provides for instruction-level disassembly." IEEE Consumer Electronics Magazine 6(3): 92-102.

[9] J. Park, X. Xu, Y. Jin, D. Forte, and M. Tehranipoor, "Power-based Side-Channel Instruction-level Disassembler." 2018 55th ACM/ESDA/IEEE Design Automation Conference (DAC), 2018, doi: 10.1109/DAC.2018.8465848.

[10] D. Krishnankutty, Z. Li, R. Robucci, N. Banerjee and C. Patel. "Instruction Sequence Identification and Disassembly Using Power Supply Side-Channel Analysis." IEEE Transactions on Computers, vol 69:11, 2020, doi: 10.1109/TC.2020.3018092.

[11] R. Gwinn, M. Matties, and A. Rubin. (2021). "Wavelet Selection and Employment for Side-Channel Disassembly." arXiv preprint arXiv:2107.11870.

[12] *ELEGOO UNO R3 Board With USB Cable Compatible With Arduino IDE*, ELEGOO Inc., 2021. Accessed on: Aug. 1, 2021. [Online]. Available: https://www.elegoo.com/products/elegoo-uno-r3-board

[13] *ATMEGA328P*, Microchip Technology Inc., 2021. Accessed on: Aug. 1, 2021. [Online]. Available: https://www.microchip.com/en-us/product/ATmega328P

[14] *AVR® Instruction Set Manual*, DS40002198A, Microchip Technology Inc., 2020. [Online]. Available: http://ww1.microchip.com/downloads/en/DeviceDoc/AVR-Instruction-Set-Manual-DS40002198A.pdf.

[15] M. Abadi, A. Agarwal, P. Barham, E. Brevdo, Z. Chen, C. Citro, et al. "TensorFlow: Large-scale machine learning on heterogeneous systems," 2015. Software available from tensorflow.org.

[16] F. Chollet and others. "Keras," 2015. Software available from https://keras.io.